\documentclass{mn2e}
\input psfig.sty

\title[AG Draconis]
      {On the resolved radio emission from AG Draconis: evidence for jet ejection?}

\author[J.\ Miko{\l}ajewska]
       {J.\ Miko{\l}ajewska
        \vspace*{1mm}\\
        N. Copernicus Astronomical Center, Bartycka  18, PL-00716 Warsaw, Poland}

\date{\fbox{\sc Draft Version}}

\pagerange{\pageref{firstpage}--\pageref{lastpage}}
\pubyear{2002}

\begin{document}

\maketitle

\label{firstpage}

\begin{abstract} The recent detection of resolved radio emission from  AG Dra by MERLIN
reported by Ogley et al. is discussed in the context of the wind environment and the 
physical parameters and geometry of this symbiotic binary system. In particular, it is 
shown that the two radio components are closely aligned with the binary axis, and their 
separation suggests their origin in jets ejected from AG Dra during the recent 
1995--98 series of oubtbursts. \end{abstract}


\section{Introduction}

AG Dra is the well-studied symbiotic binary consisting of a high-velocity, metal-poor, 
bright K-type giant and a hot white dwarf companion (e.g. Miko{\l}ajewska et al. 1995; 
Smith et al. 1996). The binary system has an orbital period of 549 days and a well-defined 
spectroscopic orbit (Fekel et al. 2001, and references therein). AG Dra is also among the 
most active symbiotic stars. Its optical light curve is characterized by a series of 
active (outbursts) and quiescent phases (e.g. Fig. 7 of G{\'a}lis et al. 1999). Although 
the activity of AG Dra, and other classical symbiotic stars is still poorly understood, 
multifrequency observations covering a few whole activity cycles indicate that in AG Dra 
this activity is related to changes of both radius and temperature of the hot component 
(e.g. Miko{\l}ajewska et al. 1995; Greiner et al. 1997). AG Dra is also one of 10 known 
galactic supersoft X-ray sources (Greiner et al. 1997). 

Recently, the star has been searched, together with all the northern supersoft X-ray 
sources, for radio emission at 5 and 8.4 GHz (Ogley et al. 2002, hereafer O02). The 
observations by MERLIN telescope have confirmed previous VLA detections. Moreover, the 
source has been resolved at the milliarcsec scale into two components of nearly equal 
brightness, and combined flux of $\sim 1$ mJy. O02 have also studied possible 
interpretations of this emission in terms of a wind environment from either the cool giant 
or the hot white dwarf. They have concluded that all their scenarios give the radio 
emission fluxes an order of amplitude lower than the observed value. 

In the following we reanalyze a possible origin of the resolved radio
emission from various wind environment, and show that it presumably arises
from jet(s) ejected from the hot component 
during its recent series of outbursts.

\section{Radio emission from AG Dra}

\subsection{Spherically symmetric steady outflow}

We point out here that  O02 have adopted unreasonably low values for both the cool 
giant wind and the hot component luminosity in their analysis. Below in this section, we 
provide a critical analysis of their assumptions and calculations.

The wind environment of AG Dra has been discussed in several papers, and estimates for the 
mass loss from both components are available. All these estimates have been based on the 
assumption of an isothermal, spherically symmetric steady flow. 
In particular, the mm-submm radio spectrum and  $K-[12]$ colour excess are both consistent 
with the cool giant mass-loss rate, $\dot{M}_{\rm g}/v \sim 2.5 \times 10^{-8}\, 
(d/2.5\,\rm kpc)^{3/2}\, \rm 
M_{\sun}\,yr^{- 1}/(km\,s^{- 1})$ (Miko{\l}ajewska, Ivison \& Omont  2002; Kenyon 1988), 
whereas the UV continuum and emission line analysis (M{\"u}rset et al. 1991) gave
$\dot{M}_{\rm g}/v \ga 4 \times 10^{-8}\, (d/2.5\,\rm kpc)\, \rm M_{\sun}\,yr^{-1}/
(km\,s^{-1})$,  in all cases values two orders of magnitude {\it larger} than 
$\dot{M}_{\rm g}/v \sim 5 \times10^{-10}\, \rm M_{\sun}\,yr^{- 1}/(km\,s^{-1})$ adopted by 
O02. Using the Wright \& Barlow (1975) formula for completely ionized wind, the first  
estimate above gives a flux at 5 GHz of $\sim 0.5$ mJy {\it independent} of the adopted 
distance. Thus, the cool component wind could, in principle, account for the observed 
intensity of the radio emission but not for the double structure. 

We also find an error in the value of irradiation luminosity adopted by O02 in their 
estimate for the mass loss rate in the cool giant due to irradiation by its hot companion, 
$L_{\rm h} = 1.4 \times 10^{36}\, \rm erg\,s^{-1}$ (or $L_{30} = 0.14$ in units of 
$10^{30}\, \rm J\,s^{-1}$; their Eq. (3) and Table 2). This error apparently stems from a 
mistake in rescaling results of Greiner (2000) to $d=1.7\, \rm kpc$ used by O02. The 
rescaling gives in fact $L_{30} = 0.44$. Using Eq.(3) of O02 with more realistic values, 
namely $L_{30} \sim 1$ (e.g. Greiner et al. 1997), $r_2\sim 80\, \rm R_{\sun}$, and $d\sim 
2.5$ kpc\footnote{Parenthetically, we also point out that the values of $d=1.7\, \rm kpc$ 
and $r_2=30\, \rm R_{\sun}$ adopted by O02 from Tomov, Tomova \& Ivanova (2000) are mutually inconsistent given the observed near IR magnitudes, colours and spectral classification of the giant. Namely, if we accept $r_2=30\, \rm R_{\sun}$, we obtain (using the Barnes-Evans relation, Cahn 1980, with the surface brightness appropriate for a mid K-type giant and the observed  $K=6.2$, e.g. Belczy{\'n}ski et al. 2000) $d\approx 1$ kpc, even more discrepant with our preferred value of $d=2.5\, \rm kpc$.} (Miko{\l}ajewska et al. 1995),  we find the mass loss rate for the evaporated wind, $\dot{M}_{\rm g} \sim 1.2 \times 10^{-7}\, \rm M_{\sun}\,yr^{-1}$, and the predicted quiescent 5 GHz flux of $\sim 0.1\, (d/2.5\, \rm kpc)^{-2/3}$ mJy. This radio flux can be further increased by increasing the hot component luminosity (e.g. Miko{\l}ajewska et al. (1995) found an order of magnitude increase in $L_{\rm h}$ due to activity), however, this model still does not account for the double source structure.

Concluding this section, the resolved radio emission from AG Dra detected by 
MERLIN cannot be interpreted in terms of the spherically symmetric steady wind 
from the cool giant. We note, however, that the {\it Hipparcos} position of AG 
Dra (Perryman et al. 1997) practically coincides with that of the  N1 component 
in the MERLIN image. This indicates that the N1 component can, in principle, 
originate in the cool giant wind.

\subsection{Episodic flow}

\begin{table} 
\begin{center} 
\caption{History of radio observations of AG Dra} \bigskip 
\begin{tabular}{@{}lccc@{}} 
\hline \smallskip 
JD & Frequency & Flux & Reference \\ 
2400000+ & [GHz] & [mJy] & \\ 
\hline 
45006 & 4.9 & $<0.41$ & ST90 \\ 
46515 & 4.9 & $0.60\pm0.21$ & ST90\\ 
46611 & 4.9 & $\ga 0.5$ & TC87 \\ 
46646 & 4.9 & $0.36\pm0.08$ & ST90\\ 
46646 & 14.9 & $0.77\pm0.23$ & ST90\\ 
48290 & 8.3 & $<0.17$ & SKT93 \\ 51622-30 & 5.0 & $\sim 1$ & O02\\ 
\hline \end{tabular} 
\end{center} \smallskip {\footnotesize ST90 -- 
Seaquist \& Taylor 1990; SKT93 -- Seaquist et al. 1993; TC87 -- Torbett \& Campbell 1987.} 
\end{table}

The active phases of AG Dra are characterized by an increase of the emission line widths 
and marked P Cyg structure of high ionization UV lines (e.g. N\,{\sc v}), which indicate 
outflow velocities of $200$--$300\, \rm km\,s^{-1}$ (Miko{\l}ajewska et al. 1995, and 
references therein), while the optical He\,{\sc ii} and H\,{\sc i} Balmer lines develop 
broad emission wings (e.g. Tomova \& Tomov 1999). This behaviour suggests that the hot 
component develops a significant wind in outburst. Tomova \& Tomov (1999) estimated the 
wind velocity of $\sim 800\, \rm km\,s^{-1}$ and  the mass loss rate of $\sim 2 \times 
10^{-7}\, \rm M_{\sun}\,yr^{-1}$ from the broad wings of H$\alpha$ and H$\beta$ profiles 
observed during the 1995 outburst. Their estimate made use of the distance-dependent 
estimate for the hot component radius of Miko{\l}ajewska et al. (1995), and thus it 
corresponds to $d=2.5$ kpc. It was also based on the assumption of the spherically 
symmetric steady wind which is probably not true for the hot component wind developed 
during the outburst. Anyway, applying the Wright \& Barlow formula (1975), such a wind 
should give rise to the radio flux of $\sim 5\, \rm \mu Jy$ at 5 GHz, again much lower 
than the flux observed by O02.

Although only a few attempts have been made to observe AG Dra at radio wavelenghts, it 
seems that the radio emission may be variable, and  possibly related to the hot component 
activity (Table 1). AG Dra was not detected at 4.9 GHz on JD 2\,445\,006 (Seaquist \& 
Taylor 1990) when the star was at maximum of a large eruption (see the light curves in 
Miko{\l}ajewska et al. 1995, and G{\'a}lis et al. 1999), whereas it was detected at 3 
epochs (Torbett \& Campbell 1987; Seaguist \& Taylor 1990) during less pronounced burst 
and its decline a few years later. It is particularly important that AG Dra was not 
detected on JD 2\,448\,290, when the star was in quiescent phase, by the highly sensitive 
VLA survey at 8.3 GHz (Seaquist et al. 1993), and the $3\,\sigma$ upper limit for the 8.3 
GHz flux of $ \la 0.17\, \rm mJy$ was much lower than the 5 GHz fluxes of any of the 
two radio components detected by O02.

The hot component entered another large outburst in 1995, which was followed by a series 
of more or less prominent bursts of activity till at least the end of 1999. It is 
interesting that there are two kinds of outbursts: stronger, cool outbursts during which 
the hot component temperature decreases, and fainter, hot outbursts during which it 
increases (Gonz{\'a}lez-Riestra et al. 1999; Miko{\l}ajewska et al. 1995). The 
temperature decrease during the strong cool outbursts is very likely due to slow expansion 
of the hot component to about 2--3 times its original size (Miko{\l}ajewska et al. 1995; 
Greiner et al. 1997) and possible development of an optically thick wind. All the previous 
radio detections were made during hot outbursts. Moreover, in 1997 and 1998, AG Dra was 
detected at 1.3 and 0.85 mm by the IRAM and JMCT telescopes, respectively (Miko{\l}ajewska 
et al. 2002). At the time of both detections AG Dra was at the hot activity stage. 

The fact that the radio emission is resolved into 
two compact components raises an interesting possibility 
that it originates in jets ejected from AG Dra.
Bipolar jets have been detected in a few symbiotic systems,
e.g. in V694\,Mon, CH Cyg, Hen\,3-1341 and StHA\,190 
(Belczy{\'n}ski et al. 2000, and references therein; Tomov, Munari \& Maresse
2000; Munari et al. 2001), and in all cases they have been associated with the hot
component outbursts and an appearance of a high-velocity stellar wind. 
A relation between the radio emission and the hot component activity stage has been also 
found in another symbiotic binary, CI Cyg. The cm-submm spectrum of CI Cyg is inconsistent 
with predictions of any model involving spherically symmetric wind(s), and the radio 
emission most likely originates from a bipolar outflow (Miko{\l}ajewska \& Ivison 2001).


\begin{figure}[t!] \centerline{\psfig{file=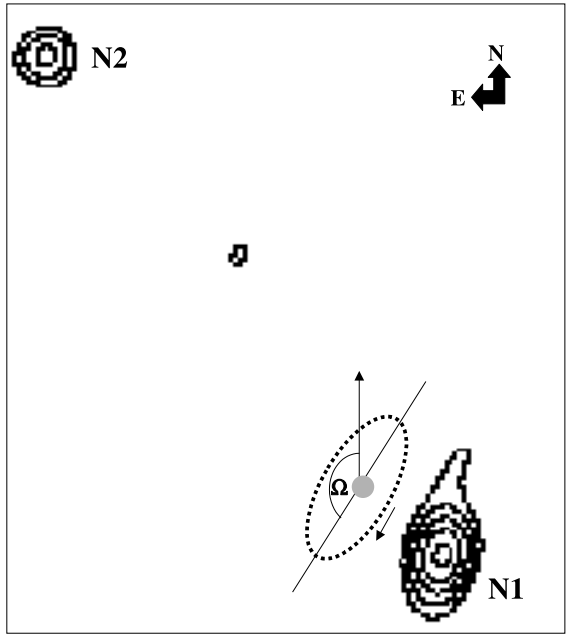,width=7.5cm,angle=-90}} 
\caption{Schematic illustration of the relationship between the binary orbit, derived from 
spectropolarimetric observations of  the $\lambda\,6825$ feature by Schmid \& Schild 
(1997), and the N1 and N2 features of  the resolved MERLIN image of AG Dra at 4.994 GHz in 
2000 March (O02). The orbit of the hot component relative to the cool giant (marked as 
the grey disk) is given by dots  (not to be scaled relative to the MERLIN image). The 
orientation of the orbital plane is $\Omega \approx 150^{\circ}$, and the system 
inclination is $i \approx 120^{\circ}$. \label{schematic}} \end{figure}

Figure 1 shows the  relationship between the binary geometry and the resolved MERLIN image 
of AG Dra. From the data in Table 1 of O02, we have estimated a separation of the 
components, $\alpha \sim 0.4 \pm 0.1$ arcsec, and a position angle of the whole structure, 
$\rm PA_{rad} \approx 35 \pm 20^{\circ}$. From spectropolarimetry we know the binary orbit 
orientation, $\Omega = 150 \pm 20^{\circ}$, as well as the orbit inclination, $i=120\pm 
20^{\circ}$ (Schmid \& Schild 1997). Thus the two radio components are within the 
observational errors aligned with the binary axis ($\Omega - {\rm PA_{rad}} \approx 115 
\pm 30^{\circ}$; Fig.~1). Moreover, the possible extended emission reported by Torbett \& 
Campbell (1987) has similar orientation. The separation of the two radio components 
resolved by MERLIN corresponds to $\Delta s \sim 1000\, [d/2.5\,\rm kpc]\, \rm a.u.$ 
Assuming that they have been ejected perpendicularly to the orbital plane with a velocity 
of the order of the escape velocity from the hot component, $v_{\rm e} = (2 G M_{\rm 
h}/R_{\rm h})^{1/2} \sim 800\, \rm km\,s^{-1}$ for $M_{\rm h} \sim 0.4$--$0.6\, \rm 
M_{\sun}$,  and $R_{\rm h} \sim 0.22\, \rm R_{\sun}$ at  the maximum of the large otbursts 
(Miko{\l}ajewska et al. 1995; Greiner et al. 1997), this separation implies that the 
ejection took place $\Delta t \sim \Delta s/2 v_{\rm e} \cos i \sim 3$ years before the 
MERLIN observations, possibly during the recent 1995--98 series of outbursts.

\section{Conclusions}

The major results and conclusions of this paper can be summarised as follows:

\begin{description}

\item ({i})  The radio emission from the symbiotic binary AG Dra seems to be variable, 
and probably related to the hot component activity. 

\item ({ii}) The cool component wind can, in principle, account for the 
intensity of the N1 component (O02), which position practically coincides with 
the {\it Hipparcos} position of AG Dra.

\item ({iii}) The two radio components resolved by MERLIN (O02) are practically aligned 
with the binary axis of AG Dra. The possible extended radio emission reported by Torbett 
\& Campbell (1987) has similar orientation. The resolved radio emission presumably 
originates in jets ejected from the binary system.

\item ({iv}) Assuming that the jet velocity is of order of the escape velocity of the hot 
component, the separation between the two radio sources indicates that the ejection took 
place $\sim 3$ yr earlier, and it was associated with the recent series of outbursts.

\end{description}

\subsection*{ACKNOWLEDGEMENTS} 
I gratefully acknowledge very helpful comments on this 
project by M. Friedjung, A. Omont, and R. Viotti. This study was supported in part by the 
KBN Research Grant No 5P03D\,019\,20, and by the JUMELAGE program "Astronomie 
France-Pologne" of CNRS/PAN.

\label{lastpage}

\end{document}